# Signs and Signatures of Intelligence

**Julia DeMarines**

ORCID: 0000-0002-7304-8481

1 Blue Marble Space Institute of Science, 600 1st Avenue, 1st Floor, Seattle, WA 98104, USA

2 Department of Earth & Planetary Science, University of California, Berkeley, 307 McCone Hall, Berkeley, CA 94720, USA

3 The Decentralised Human Architecture Foundation, Level 5/171 La Trobe St, Melbourne VIC 3000, Australia

## Abstract

Research into the emergence and evolution of intelligence is underrepresented in astrobiology. I propose noosemiotics, the study of noosignatures, to address this gap. Noosignatures are the structured traces that minds leave on a medium, whether physical (tools, etchings, architecture) or signal-based (encoded transmissions, complex animal communication), that remain detectable as the products of intelligence regardless of whether their meaning can be recovered. Where biosignatures mark life and technosignatures mark technology, noosignatures mark the space between. This paper defines noosemiotics, situates it within astrobiology, and proposes a research program for empirical metrics, with material culture as the primary test bed.

## 1. Introduction: The Gap in Astrobiology

The search for life in the universe is, at its core, a search for anomalous signatures distinguishable from natural phenomena and known astrophysical processes. NASA defines astrobiology as "the study of the origin, evolution, distribution, and future of life in the universe," and the field encompasses a wide range of expertise including the physical sciences and the humanities. In the physical sciences the range includes prebiotic chemistry and extremophile biology to planetary habitability and the search for technologically advanced civilizations. Within this breadth, two observational subfields have emerged as primary research programs. Biosignature research investigates the observable products of life, including atmospheric gases such as $O_2$, $O_3$, and $CH_4$ that, in combination, suggest biological activity. Technosignature research investigates the observable products of technology. Examples include, but are not limited to, electromagnetic signals, waste heat, atmospheric pollutants, and large-scale structures that imply a civilization capable of planetary-scale engineering.

Between these two programs lies an underexplored domain. The evolutionary trajectory from the first living cells to a species capable of broadcasting radio signals spans roughly 3.5 billion years and includes the emergence of cognition, cumulative culture, language, and complex social organization. These are not incidental steps. They are necessary preconditions for any civilization that produces technosignatures, at least on our planet. Yet the study of intelligence as a detectable, observable phenomenon has received comparatively little formal attention within astrobiology.

This disparity is empirically visible. At the 2026 Astrobiology Science Conference (AbSciCon), the largest astrobiology gathering in the world, biosignature research commanded 23 dedicated sessions and an estimated 300 to 450 submitted abstracts. Technosignature and SETI observational research accounted for one session and 14 abstracts. Intelligence research as a conceptual and empirical program claimed zero dedicated sessions and a single abstract (Table 1). The imbalance reflects scientific priorities and the absence of an established framework for studying intelligence as something that leaves physically detectable traces.

| Abstract Category | Dedicated sessions | Submitted abstracts |
|---|---|---|
| Biosignatures + detection methods | 23 | est. ~300-450 |
| Technosignatures / SETI (observational) | 1 | 14 |
| Study of intelligence (conceptual) | 0 | 1 |

*Table 1: Session counts queried from the 2026 AbSciCon program are exact. Abstract counts: technosignature (14) and intelligence (1) are exact, hand-counted from full session rosters; the biosignature abstract figure is an order-of-magnitude estimate pending a session-by-session program pass.*

Part of the explanation lies in how the field has defined its terms. A 2018 "ad-hoc nomenclature committee" redefined 'intelligence' in the SETI context as the capacity to engineer technology detectable by astronomical means, collapsing the study of intelligence into the study of technosignatures (Wright et al., 2018; Tarter, 2007). This is a useful operational move for observational programs and aligns with the colloquial understanding of what SETI does in practice, but it excludes the question of what intelligence leaves behind before a civilization produces interstellar-range technology, and what traces emerge and accumulate across the long interval between the origin of life and the advent of radio.

There is growing recognition that this interval matters scientifically. The fraction of life-bearing worlds that develop intelligence (fi in the Drake Equation) remains the least constrained of the equation's biological terms, in part because no empirical framework exists for detecting or measuring intelligence in matter or in signals. Duner (2017) identified this gap from a cognitive-semiotic perspective, noting that fi, fc, and L rest on unexamined assumptions about what intelligence and communication actually are, and calling for a field he termed astrocognition to address it.

Here I propose noosemiotics as a complementary empirical framework. Noosemiotics is the study of noosignatures: the structured traces that minds leave on a medium, whether physical or signal-based, that remain detectable as the products of intelligence regardless of whether their meaning can be recovered. The sections that follow define the framework, situate it within astrobiology's existing structure, and propose a research framework built on complexity metrics applicable to the material and signal record of cognitive evolution.

## 2. Biosignatures, Technosignatures, and the Space Between

Earth's own evolutionary record offers our only model for understanding the temporal relationship between signatures of life and signatures of technology. The planet formed approximately 4.5 billion years ago. Microbial life emerged within the first billion years, and biological activity became atmospherically detectable through methane produced by methanogenesis roughly 3.5 billion years ago. Life's detectability rests not on any single molecule but on the chemical disequilibrium that biological activity imposes on a planetary atmosphere: living systems continuously produce reactive gases that would not otherwise coexist, and it is the departure from thermodynamic equilibrium that signals their presence (Lovelock, 1965; Sagan et al., 1993). The Great Oxidation Event, approximately 2.4 billion years ago, introduced oxygen as the most widely studied atmospheric biosignature candidate, most powerfully when detected alongside methane. A broader suite of atmospheric biosignatures has since been identified, including nitrous oxide, dimethyl sulfide, and the vegetation red edge, each produced by distinct biological processes and detectable under different observational conditions (Schwieterman et al., 2018). For the following two billion years, the observable atmospheric record changed relatively slowly, reflecting the long dominance of microbial life and, later, the gradual rise of complex multicellular organisms. Radio transmissions represent Earth's first widely recognized technosignature, with high-power signals potentially detectable from nearby interstellar distances for approximately one hundred years.

An observer monitoring Earth remotely across this timeline would have witnessed the emergence of atmospheric biosignatures billions of years ago, followed by a long interval of modest change, followed by an abrupt and unmistakable technological signal in the twentieth century. What that observer would not see is the 3.5 billion years of cognitive and cultural evolution that made the technological signal possible: the emergence of nervous systems, the development of social cognition, the accumulation of tool-making traditions across generations, the appearance of symbolic behavior, and the rise of language and writing. These transitions are not absent in the physical record. They are simply not yet part of any systematic search framework.

The nitrogen record provides a concrete illustration. Kopacek and Posch (2011) reconstructed anthropogenic reactive nitrogen emissions across the entire Holocene and found that 70 to 84 percent of cumulative human nitrogen emissions occurred before industrialization, beginning with the advent of agriculture approximately 8,000 years ago. Additionally, Haqq-Misra et al. (2022) identified nitrogen cycle disruption as a candidate technosignature detectable in exoplanet atmospheres, though their model requires a planetary population of 30 to 100 billion to produce detectable signal levels. Together, these findings suggest that the material signature of a developing intelligence may be legible in planetary geochemistry thousands of years before that intelligence produces radio signals, and potentially in the deep paleorecord of a planet's own surface long after those signals have faded.

Sheikh et al. (2025) demonstrated that Earth's current technosignatures span thirteen orders of magnitude in detectability, with intermittent planetary radar dominating. This range illustrates how unevenly the noosphere writes itself into the observable record. The biosignature-to-technosignature interval on Earth is not empty; it is populated by traces of increasing cognitive and material complexity that no existing search framework is designed to detect.

The Drake Equation term that corresponds to this interval is fi, the fraction of life-bearing worlds that develop intelligence. It remains the least empirically constrained of the equation's biological parameters, in part because no singular metric exists for measuring intelligence as a physical phenomenon. Wright et al. (2022) argued that technosignatures may be more abundant, longer-lived, and less ambiguous than biosignatures.

Noosignatures extend that argument further back in time: the traces of intelligence may predate the traces of technology by millions of years. On worlds where the noospheric transition began but never produced interstellar-range technology, noosignatures may be the only traces available.

## 3. Defining Noosemiotics: Signs, Signatures, and the Medium

The space between biosignatures and technosignatures is not merely a gap in our search strategies; it reflects a gap in conceptual vocabulary. What it lacks is a precise framework for describing physically detectable traces of intelligence that do not rise to the threshold, or off-planet detectability, of interstellar technology. Noosemiotics is proposed as that framework. Its fundamental unit is the noosignature: a structured trace left on a medium by a mind, where the trace remains detectable as the product of intelligence regardless of whether its original meaning can be recovered.

A precise definition requires a distinction that is easy to overlook. In semiotics, a sign is meaning-laden, culturally transmitted, and interpretant-dependent: it signifies something, like hospitality, danger, the sacred, a grammatical rule, but only to a mind equipped with the cultural context to read it. A signature, by contrast, is causally embedded, physically detectable, and interpretant-independent. It remains traceable to the history that produced it whether or not any interpreter survives to read it. For example, a deciduous tree's age is recorded in its rings whether or not anyone counts them. A molecule's bond history is embedded in its structure whether or not anyone runs a spectrometer across it. A noosignature is the signature left specifically by a mind: the physical residue of causal, cognitive activity encoded into a medium (Wolpert, 2003; Walker & Davies, 2013).

The distinction has practical consequences. Meaning decays as the culture that sustained it disperses. The Indus Valley script has not been deciphered; the ritual purpose of many Paleolithic ochre marks may never be recovered. But the structural complexity of the objects that carried that meaning does not decay in the same way. A worked flint is identifiable as worked long after any account of who made it or why has been lost. A hearth is a hearth by its chemical and structural properties, not by the social meaning the fire once held. Archaeology already practices this: it reconstructs intelligence from physical traces whose meaning is unrecoverable, by reading structural complexity that no natural abiotic process could have produced. Noosemiotics generalizes this practice across scales and media, asking what framework can make it systematic and, eventually, detectable beyond Earth.

The medium of a noosignature may be physical or signal-based. Physical media include tools, structures, cultivated landscapes, chemical modifications to soil or atmosphere, and any material object whose structural complexity exceeds what stochastic processes could generate. Signal-based media include any patterned transmission whose organizational complexity implies a generating mind: the combinatorial structure of whale song (Arnon et al., 2025), the recursive syntax of human language, the information-dense encoding of electromagnetic transmissions.

A clarification is necessary for wave-based signals specifically. Acoustic communications such as whale song travel through a medium rather than leaving a trace in one: sound propagates through water or air and dissipates without embedding its structure in a persistent substrate. As Landauer (1991) established, information is physical and must be instantiated in matter to persist. A whale song, or similarly an alien radio transmission, that goes unrecorded leaves no durable noosignature; only when captured by a technological recording device, such as a hydrophone paired with an analog-to-digital converter, does the wave-based signal become a physical trace available for analysis. For the purposes of noosemiotics, the signal-based noosignature is not the transmission itself but the physical record of the transmission, cast as a 'causal shadow'. Electromagnetic signals occupy a different category: propagating through space at the speed of light, they carry their structure across interstellar distances without requiring an intermediate recording medium. The breadth of this definition is intentional. Noosemiotics does not begin with radio and work backward; it begins with the minimal conditions for a physically detectable trace of cognitive agency and asks how those conditions are met across different substrates, timescales, and levels of cognitive development (Duner, 2017).

Noosignatures are therefore distinct from both biosignatures and technosignatures in a precise sense. Biosignatures mark the presence of life; they do not require cognitive agency. Technosignatures mark the presence of interstellar-range technology; they set the threshold too high to capture the long developmental interval that precedes them. Noosignatures occupy the interval between: detectable in principle wherever intelligence has structured a medium, whether that intelligence produced radio transmissions or only stone tools. This is the domain noosemiotics proposes to study.

## 4. A Physics of Noosignatures: Toward Empirical Metrics

A noosignature framework is only useful if the traces it describes can be measured. The sign/signature distinction offered in the previous section is conceptually precise, but precision is not measurement. What does it mean, in practice, to detect the structural complexity that implies a generating mind? This section outlines a candidate answer.

One approach begins with non-equilibrium thermodynamics. Living systems maintain local order by processing energy gradients and exporting entropy to their environment, a foundational insight connecting physics to biology (Schrodinger, 1948). Kondepudi (2012), building on Prigogine's theory of dissipative structures, formalized how self-organizing systems arise and persist far from thermodynamic equilibrium by continuously processing energy flows, and proposed that physical intelligence is an emergent property of this self-organizing capacity. Intelligence extends this ordering in a specific direction: Wolpert (2003) identified causal belief, the cognitive capacity to model cause and effect, as the property distinguishing human tool use from the behavior of other primates, arguing that a complex tool cannot be made without a mind that understands causation. Wissner-Gross and Freer (2013) proposed a complementary formal model in which intelligence corresponds to the tendency of adaptive agents to preserve future causal options. These perspectives converge on a shared implication: minds produce structured complexity of a kind that physical processes alone cannot generate.

No consensus metric yet exists across these approaches. Assembly Theory offers a tractable empirical bridge. The assembly index of an object is defined as the minimum number of joining operations required to construct it from its elemental components (Sharma et al., 2023; Cronin and Walker, 2026). The key property is that objects above a threshold assembly index cannot arise by chance: their construction requires memory of prior steps, which in practice means a system capable of storing and transmitting causal information across time. The assembly index is therefore not simply a measure of complexity, but a measure of the causal history embedded in physical structure. This is precisely the invariant property that defines a noosignature in the previous section: not meaning, but the physical record that a mind was here. Mastrogiovanni (2025) has noted the bridge to semiotic theory directly: AT's property of iterability, the requirement that high-assembly objects be constructable step-by-step and each step be reproducible, is the physical analog of the sign vehicle in semiotics, meaning the carrier can be detached from its context and reproduced across time and culture.

Precedents for applying complexity metrics to historical processes exist: Delahaye and Vidal (2018) proposed logical depth as a measure of organized complexity in big history, situating the rise of complexity across cosmic and cultural timescales within a computational framework, and anticipating the kind of quantitative trajectory that an Assembly Theory approach to the noosignature record would produce.

Assembly Theory was developed as a biochemistry-agnostic life-detection framework, demonstrated on organic molecules and applied to distinguish biological from abiotic chemistry (Kahana et al., 2024). Patarroyo et al. (2025) extended it to inorganic materials, showing that crystals and engineered microprocessors display distinct assembly signatures and that the index scales with the degree of directed manufacturing. Material culture artifacts, from worked stone to Bronze Age implements to writing systems, are the natural next test bed. The cognitive engine that drives rising complexity across this record has been characterized as cumulative technological culture: the species-specific capacity for technical reasoning that allows each generation to improve on the tools it inherits, translating causal belief at the individual level into rising assembly complexity at the population level over generational time (Osiurak and Reynaud, 2020). Applied to a temporal series spanning the archaeological record, AT-style metrics should produce a complexity curve that tracks the emergence and development of the noosphere as a measurable physical phenomenon. This is the empirical program noosemiotics motivates.

For signal-based media, complementary metrics are available. Mutual information content, syntactic combinatoriality, and recursive structure have been used to probe the complexity of animal communication systems, including cetacean song and primate vocalizations (Duner, 2017). These are not interchangeable with the assembly index, which measures physical construction steps rather than informational structure, but they are consistent with the same underlying intuition: that minds produce signals whose organizational complexity exceeds what stochastic generative processes can account for.

I note intentionally that the metrics outlined here are proposed, not yet applied to material culture in a systematic way. What this paper establishes is the conceptual and theoretical architecture that makes such a research program coherent. Whether a rising AT complexity curve is detectable in the deep archaeological record, and whether it is sufficiently distinctive to serve as an exoplanetary detection target, are empirical questions the proposed program is designed to address.

## 5. Case Studies: The Noosignature Record on Earth

Earth's own archaeological record is the most immediate test bed for noosemiotics, and it offers a striking span of material to work with. From the earliest known stone tools to the emergence of writing systems, the terrestrial record should, on this framework's prediction, display a rising assembly complexity curve that tracks the development of the noosphere as a physical phenomenon. Paige and Perreault (2024) provide empirical evidence of precisely this trajectory, quantifying stone tool complexity across 3.3 million years of the archaeological record and documenting a marked rise in the upper limit of technological complexity during the Middle Pleistocene. This section sketches the proposed empirical program rather than reporting completed results; the analysis remains future work and is presented here to demonstrate the testability of the framework.

The earliest known stone tools, the Lomekwian assemblage from Lomekwi 3 in Kenya, date to approximately 3.3 million years ago and represent the oldest physical evidence of intentional stone modification by hominins. The Oldowan lithic assemblage, appearing around 2.6 million years ago, marks a further threshold: the conchoidal fracture patterns on Oldowan cores require controlled, directional force applied with an understanding of how stone breaks, and are not reproducible by natural percussion or non-human primates under experimental conditions. By any reasonable complexity threshold, these objects sit above the background assembly distribution of unmodified rock. The Acheulean biface, appearing around 1.8 million years ago, pushes further still: the symmetric, standardized form implies deliberate construction and, novelly, the transmission and replication of a template across generations. This marks the earliest recoverable signature of cumulative technological culture (Osiurak and Reynaud, 2020; Paige and Perreault, 2024). The recurring geometric forms of Acheulean tools across geographically separated populations suggest that cognitive constraints on what kinds of complexity get produced and transmitted are themselves cross-culturally stable (Boyer, 1998).

The symbolic record continues to record informational offloading (Donald, 1991). Ochre processing, perforated beads, and geometric petroglyphs, appearing between roughly 500,000 and 100,000 years ago depending on site, mark the emergence of objects whose assembly complexity exceeds functional necessity (Henshilwood et al., 2011). They encode information beyond use, a process Clark and Chalmers (1998) would recognise as the extension of cognitive processes into the environment. The process-intensive foods of the Neolithic, fermented beverages, leavened bread, and cooked legumes, represent a different category of noosignature: multi-step chemical transformations requiring accumulated procedural knowledge, their assembly index detectable in residue chemistry long after the cultural context that produced them has been lost (Arranz-Otaegui et al., 2018; McGovern et al., 2004).

Writing, emerging around 5,000 years ago, marks a local peak, or turning point, of informational offloading: recursive, iterable, and combinatorial, it is the material form in which the noosphere made its own structure legible. Spier (2011) identifies writing as the onset of a higher level of information complexity in Big History, qualitatively distinct from genetic and cultural transmission (Donald, 1991; Spier, 2011).

The complexity curve does not terminate with writing. Artificial intelligence systems (AI and LLMs) represent a new class of noosignature best described as post-biological: produced not directly by biological minds but by systems that are themselves compressed records of human cognitive output. Large language models and other AI architectures are trained on the accumulated textual and symbolic production of human civilization; the causal history of millions of minds is embedded in their parameters and reflected in their outputs. Under the framework's own logic, the biological origin of that causal history is upstream rather than absent. Clark and Chalmers (1998) established the precedent that cognitive processes can extend beyond the biological organism into tools and environments; AI represents the most extensive and recursive instantiation of this principle yet produced, an external system that stores and recombines the noosphere's signal-based record at civilizational scale. At the network level, emergent patterns in distributed AI behavior and human-AI interaction generate noosignature-class complexity that no individual mind authored, raising new questions about collective cognitive agency at the planetary scale (Frank, Grinspoon and Walker, 2022). Whether AI outputs are classified as noosignatures or as a subcategory of technosignature will depend on context: as artifacts, they record causal cognitive history and are noosignatures; as interstellar transmissions, they would qualify as both. What is clear is that any adequate noosemiotic framework must be able to account for intelligence that has partially migrated from its biological substrate, and that the emergence of AI accelerates the urgency of developing that framework.

The archaeological noosignature record is complemented by a geochemical one. Kopacek and Posch (2011) document a 10,000-year nitrogen isotope record in European lacustrine sediments showing anthropogenic disruption beginning with Neolithic agriculture, roughly 8,000 years before any detectable technosignature. This is a noosignature in planetary geochemistry rather than artifact form, and it speaks directly to Frank, Alberti, and Kleidon's (2017) concept of the hybrid planet: a world whose surface chemistry has been measurably reorganized by the activity of a living system before that system reaches industrial technology. The noospheric transition, on this evidence, begins far earlier than the Anthropocene framing suggests.

## 6. Discussion and Outlook

Noosemiotics is not an Earth-specific framework. Any planet that has developed intelligence should possess a noosignature record, a physical history of causal cognitive activity impressed into whatever media were available, if available. That record is potentially detectable before technosignatures emerge and may persist after they have faded or ceased. This is the temporal argument for taking noosemiotics seriously as a search strategy: the interval between the first noosignature and the first detectable technosignature on Earth spans millions of years, and there is no principled reason to expect that interval to be shorter elsewhere.

The cooperation barrier identified by Vidal (2026) refines the point. A civilization that develops intelligence but fails to solve the coordination problems required to sustain a cooperative planetary noosphere may produce noosignatures across geological timescales without ever achieving the technological coherence needed to generate detectable technosignatures. For such worlds, noosignatures may be the only recoverable evidence that intelligence existed at all. The distinction between fi and fc in the Drake Equation is, on this view, not simply a matter of degree; it may mark a genuine developmental threshold that some worlds cross and others do not. Noosemiotics gives us a way to look for the worlds that did not make it across this technological threshold.

Within the existing search framework, noosemiotics suggests a fifth strategy alongside the four outlined by Chyba and Hand (2005). Rather than searching for specific physical substrates, electromagnetic modalities, or engineering signatures, a noosemiotic search reads the semiotic structure and assembly complexity profile of candidate signals and surfaces, asking whether their organizational properties imply a generating mind regardless of what that mind was made of or how it communicated. Anomaly detection approaches calibrated against expected abiotic complexity distributions, of the kind developed by Singam et al. (2020), are a natural computational complement to this strategy.

I acknowledge the limitations of the framework as currently developed. Assembly Theory has not yet been applied systematically to material culture, although Patarroyo et al. (2025) apply its fundamentals to quantify complexity in inorganic molecules and solid-state periodic objects including crystals, minerals, and microprocessors. Signal-based complexity metrics remain underdeveloped as noosignature indices; and the proposed Earth case studies are outlines rather than completed analyses. These are the research program's open questions, and the open questions a noosemiotic research program is designed to address. What the framework already offers is a precise vocabulary for a phenomenon that astrobiology has so far studied only in pieces: the physically detectable, interpretant-independent trace of a mind at work on a world.